\documentclass[useAMS,usenatbib]{mn2e}
\usepackage{graphicx}
\usepackage{deluxetable}

\newcommand{\HI}{\hbox{{\rm H}{\sc \,i}}}
\newcommand{\Htwo}{\hbox{{\rm H}$_2$}}

\newcommand{\lya}{\hbox{{\rm Ly}$\alpha$}}

\newcommand{\Ha}{\hbox{{\rm H}$\alpha$}}
\newcommand{\mpy}{\hbox{M$_{\odot}$\,yr$^{-1}$}}
\newcommand{\lsun}{\hbox{L$_{\odot}$}}
\newcommand{\msun}{\hbox{M$_{\odot}$}}
 \newcommand{\rhosun}{\hbox{M$_{\odot}$\, Mpc$^{-3}$}}
\newcommand{\cmsq}{\hbox{cm$^{-2}$}}

\newcommand{\hMpcc}{\hbox{$h^3$~Mpc$^{-3}$}}
\newcommand{\hMpc}{\hbox{$h^{-1}$~Mpc}}

\newcommand{\kms}{\hbox{${\rm km\,s}^{-1}$}}

\def\bsp_small{\vspace{0.5cm}\small\noindent This paper
has been typeset from a \TeX / \LaTeX\ file prepared by the author.}

\title[The metals in SMGs]{The missing metals  problem. I. How many metals
are in submm galaxies?}
\author[N. Bouch\'e, M. D. Lehnert, C. P\'eroux]{Nicolas
Bouch\'e$^1$\thanks{E-mail: nbouche@mpe.mpg.de (NB)}, 
Matthew D. Lehnert$^1$,
C\'eline P\'eroux$^2$\\ 
$^1$Max Plank Institut f\"ur extraterrestrische Physik, Giessenbachstra\ss e, D-85748 Garching, Germany \\
$^2$European Southern   Observatory, Karl-Schwarzschild-Str 2, D-85748 Garching, Germany}  
  
\begin{document}

\date{Accepted ---.
      Received ---;
      in original form ---}

\pagerange{\pageref{firstpage}--\pageref{lastpage}}

\maketitle

\label{firstpage}

\begin{abstract}
At redshifts larger than 2, a large fraction (80\%) of the metals are
apparently yet undetected.  We use a sample of sub-mm selected galaxies
(SMGs) with molecular gas and dynamical mass measurements from the
literature to put constraints on the contribution of such galaxies
to the total metal budget.  Compared to Lyman break galaxies (LBGs),
for example, SMGs are rarer (by a factor of 10 or more), but contain
much more gas   and are more metal rich. For SMGs brighter than 3~mJy,
we estimate that SMGs contain only $\la$9\%\ of the metals when we combine the observed dynamical
masses ($<M_{\rm dyn}>\sim$few$\times 10^{11}$~\msun),   number density ($n\simeq 10^{-4}$~Mpc$^{-3}$),
  observed gas metallicity ($Z\simeq$1--2$Z_\odot$), and   observed gas fractions ($f_{\rm gas}\approx$~40\%) 
  assuming a molecular to neutral hydrogen ratio of 1.
Including SMGs fainter than 3~mJy, we estimate that SMGs contain about $\leq$15\%\ of the metals,
where our incompleteness correction is estimated from the dust mass function. 
 Our results are strong upper limits given that high gas
fractions and high overall metallicity are mutually exclusive.
  In summary, SMGs make a significant contribution to the metal budget ($\la$15\%) but not sufficient to
solve the `missing metals problem'.
A consequence of our results is that SMGs can  only add $\approx\ 3.5$\%\ to $\Omega_{\rm DLA}$,
and can not be the source of a significant population of dusty DLAs.
\end{abstract}


\section{Introduction}

A direct consequence of star-formation and, in particular, of the star-formation history (SFH)
\citep[][and others]{LillyS_96a,MadauP_96a,GiavaliscoM_04a,HopkinsA_04a}
is the production of heavy elements, known as metals. 
Indeed, for a given initial mass function (IMF), 
the total expected  amount of metals $\rho_{Z,\rm expected}$ formed by a given time $t$ is simply the
integral of the star formation density (SFD) $\dot{\rho}_{\star}(t)$ times
 $<p_z>$, where $<p_z>$ is the mean stellar yield \citep{SongailaA_90a,MadauP_96a}:
$\rho_{Z,\rm expected}=  <p_z> \int_0^t \mathrm d t \; \dot{\rho}_{\star}(t) \;. \label{eq:metalprediction}$
Using a Salpeter IMF and the type II stellar yields (for solar metallicity)
  $p_z(m)$ from \citet{WoosleyS_95a}, \citet{MadauP_96a} found that 
$<p_z>=\frac{1}{42}$ or 2.4\%. 
Using the SFH parameterized (in a LCDM cosmology) either as in \citet{ColeS_01a} or by a constant star formation rate
(SFR) beyond $z=2$, we find that the total co-moving metal density is \citep{BoucheN_05e}:
\begin{eqnarray}
\rho_{Z,\rm expected}&\simeq &4.0\times 10^6\; \rhosun\label{eq:metals:madau}  ,
\end{eqnarray}
after integrating the SFH over the redshift $z$ range from 4 to 2.
This is about 25\%\ of the $z=0$ metals.

But at redshifts $z\simeq3$, our knowledge of the cosmic metal budget
is still highly incomplete.  Indeed, only a small fraction (20\%) of
the budget is actually seen when one adds the contribution of the \lya\
forest ($N_{\HI}=10^{13-17}$~\cmsq), damped \lya\ absorbers
(DLAs) ($N_{\HI}>10^{20.3}$~\cmsq), and  galaxies such as Lyman break
galaxies (LBGs) \citep{PettiniM_99a,PagelB_02a,PettiniM_03b,WolfeA_03b}.

To account for the remaining 80\%\ of Eq.~\ref{eq:metals:madau}, or the
``missing metals,'' there are    two likely possibilities, as pointed out by
\citet{PettiniM_03b}. Either they are in a galaxy population not yet
accounted for in the budget of \citet{PettiniM_03b},
 or they are in a hot phase which is currently difficult to detect.  
In \citet[][hereafter paper~III]{BoucheN_05e}, we discuss
further the missing metal problem and the latter alternative.
In this paper and in \citet{BoucheN_05d} (paper~II), we discuss the former.
In paper~II, we discuss the contribution of both the
$z\sim2.2$ UV selected galaxies, ``BX'' \citep{SteidelC_04a}, and
near-IR selected galaxies \citep[e.g.][]{FranxM_03a}.
In this paper, we discuss submm selected galaxies (SMGs). 
SMGs are potentially good candidates
for hiding metals given that they are both gas and metal rich. For
instance, \citet[][hereafter D03]{DunneL_03a} explored the contribution
of SMGs to the metal budget using the dust mass function (DMF) of high
redshift ($z\simeq 2$--3) submm galaxies constructed from deep blank
field SCUBA surveys.  From the DMF, D03 inferred, using chemical models,
the co-moving density of metals and baryons associated with the ISM of
submm galaxies (SMGs).  They concluded that all of the remaining metals
(80\%) are in the ISM of SMGs and that the mere existence of SMGs is
enough to close the metal budget.

Recently, the gas content and metallicity of SMGs have
been directly estimated or constrained in a few cases
\citep{GenzelR_03a,NeriR_03a,GreveT_05a,TeczaM_04a,SwinbankA_04a}. These
measurements now allow us to put more direct limits on the
contribution of SMGs to the cosmic metal budget. In this paper, we will show  
 that indeed SMGs contribute
significantly to the metal budget, but their contribution is
$\la$10\%\ and, even optimistically, cannot be more than $\sim
20$\%.  In paper~II, we show that $z\sim2.2$ galaxies contribute significantly to the metal
budget, up to 15--20\%.
  Thus, combining all the know galaxy populations at $z>2.0$, there is about 50\%\
of the metal budget that has been accounted for.  In paper~III,
we will  explore whether the remaining metals have been expelled from
small galaxies into the IGM (i.e. into a hot non-detectable phase).

In section~\ref{section:dunne}, we compare our results to those
 of  D03 in a $\Lambda$CDM cosmology. 
 
In the remainder of this paper, we used  $H_0=70~h_{70}$~\kms~Mpc$^{-1}$,
$\Omega_M=0.3$ and $\Omega_\Lambda=0.7$.

 \section{How many metals in SMGs?}
\label{section:SMG:observations}

In this section,  we use the recent observations of gas content ($H_2$)
and dynamical masses of 7 SMGs \citep{GenzelR_03a,TeczaM_04a,GreveT_05a}
to put constraints on the contribution of the SMGs to the metal budget.
We first summarize the observations.

\subsection{Properties of SMGs}

\begin{table*}
\caption{Gas masses and dynamical masses of SMGs with $z>2$ and an intrinsic flux $S_{850}$ greater than 3~mJy.
The average quantities are shown.
References: (1) \citet{GenzelR_03a}, (2) \citet{NeriR_03a}, (3)\citet{GreveT_05a}, (4)   \citet{TeczaM_04a},
(5) \citet{SwinbankA_04a}.
\label{table:submm:observations}}
\begin{tabular}{lrrrrrrrr}
\hline
Name 			&   $S_{850}$~\tablenotemark{a}	
				   & 	$z_{CO}$&	$M_{\rm gas}$~\tablenotemark{b}	
				   				& 	FWHM~\tablenotemark{c}	
										&	$M_{\rm dyn}$~\tablenotemark{d} 
												&	Refs 
													&   [O/H]~\tablenotemark{e}
															& Refs \\
			&   (mJy)  &		&	$\times 10^{10}$\msun\	
								&	\kms\	&	$\times 10^{10}$\msun\		
												&	&  		&  	\\
			\hline \\
SMMJ02399$-$0136 	&  9.6 	   &    2.8076  &  $6.0$     	&  1100	     	& 60      	&  1 	&  $\cdots$   	&	  \\
SMMJ09431$+$4700 	&  8.8 	   &    3.3460 	&  $2.0$     	&   420	     	&  5/7      	&  2,3	&  $\cdots$	&	  \\
SMMJ131201$+$4242	&  6.2 	   &    3.408   &  $4.2$     	&   530         &  12     	&  3 	&  $\cdots$	&       \\
(SMMJ14011$+$1152~\tablenotemark{f}
			& 2.9      &   2.5652	&  $3.4$     	&   190	     	& $ 6$    	& 1,3	&   0.3 	&     4  )\\
SMMJ16358$+$4057 	&  8.2 	   &    2.3853  &  $5.6$    	&  840          & 9/35 		& 2,3	&  $\cdots$	&	  \\
SMMJ16366$+$4105	&  10.7	   & 	2.450   &  4.6		& 870  		& 9/37 		&  2,3  & 0.1 &  5 \\
SMMJ16371$+$4053	&  10.5	   & 	2.380   & 2.4		& 830		& 34 		&  3	& -0.1  &  5 \\
SMMJ22174$+$0015 	&  6.3 	   &    3.099   & 3 		& 780		& 28 		&  3	& $\cdots$	&  \\
 \hline  
average~\tablenotemark{g}:		& 8.6	 &		& $3.97\pm1.55$ &   767$\pm$226	& $21\pm21$/30$\pm$17 & &  &  \\
\hline
\end{tabular}

\tablenotetext{a}{Intrinsic submm fluxes.}
\tablenotetext{b}{Molecular gas masses from CO line emission.}
\tablenotetext{c}{Full width at half maximum of the CO line.}
\tablenotetext{d}{Dynamical masses for the inclination $i=45$  and using $h_{70}=1$.}
\tablenotetext{e}{[O/H] metallicities using (O/H)$_\odot$   from    \citet{AsplundM_04a}.}
\tablenotetext{f}{The amplification of this source was revised from 2.5 to $\sim$5 \citep{SmailI_05a}
moving it just below our threshold of 3~mJy.}
\tablenotetext{g}{Excluding SMM14011$+$1152.}
\end{table*}

Recent measurement of the dynamical and gas masses of currently now a dozen
$z\simeq2.5$ submm sources   have been made from CO line emission (some
resolved) using both OVRO and the IRAM Plateau de Bure interferometer
\citep[e.g.][]{FrayerD_99a,GenzelR_03a,NeriR_03a,GreveT_05a}.
They have redshifts spanning 1.0--3.3, bolometric luminosities
$L_{\rm bol}\sim 10^{13}$\lsun, and have large molecular mass  $M_{\rm
gas}\ga2\times 10^{10}$\msun\ and dynamical masses $M_{\rm dyn}\ga
0.5\times10^{11}$\msun.  From the compilation by \citet{GreveT_05a}, we
find that out of the dozen SMG with CO detections, 7 meet the following
two criteria: (i) $z>2$ and (ii) an intrinsic (de-lensed) $S_{850}$
flux $>$3~mJy.  The redshift cut-off is natural given the aims of
this paper, and the flux threshold corresponds to the one used by D03
(see section~\ref{section:dunne}).  We note that SMMJ14011$+$1152,
with an intrinsic $S_{850}$ flux of 2.9~mJy,  could be included in our
sample if the magnification is slightly smaller. Our mean values
do not change significantly if one includes SMM14011$+$1152.

The properties of these sources are listed in
Table~\ref{table:submm:observations}. One can see from the table that,
on average, $\sim$20\% (and up to 50\% for J01411) of  the dynamical
mass of SMGs is made of molecular ($H_2$) gas.  Excluding SMMJ01411,
the averaged gas mass, velocity width and dynamical masses are
$M_{\rm gas}=4.0\times 10^{10}$~\msun, FWHM$\simeq700$~\kms, $M_{\rm
dyn}\simeq2.1\times10^{11}$~\msun, respectively
(see the bottom of Table~\ref{table:submm:observations}).

Very few SMGs have had their gas phase metallicities measured or
constrained.  In the case of J14011, \citet{TeczaM_04a} used the
near-infrared integral field spectrometer, SPIFFI (now SINFONI) on the
ESO-VLT to measure the nebular emission line ratios of J14011.  Using the
classical optical diagnostic ratio, R$_{23}$(=[OII]$\lambda\lambda$3726,
3729 + [OIII]$\lambda\lambda$4959, 5007/H$\beta$), \citet{TeczaM_04a}
inferred a metallicity of $+0.27^{+0.11}_{-0.15}$dex ($Z\sim
1.9_{-0.5}^{+0.5}\;Z_\odot$).  \citet{SwinbankA_04a} used long-slit
spectroscopy to measure the [NII]/\Ha\ ratio  to infer metallicities using
the  calibration of \citet{PettiniM_04a}.  The median of their sample
is slightly below solar.
Broadly speaking, SMGs have metallicities
close to solar and up to $\sim2$ $Z_\odot$.

In order to estimate the total metal contribution of submm galaxies, it is
necessary to estimate the true number density of SMGs corrected for
the ``duty cycle,'' the fraction of cosmic
time over which submm galaxies are observed.  
 \citet{GenzelR_03a} estimated the raw co-moving density $n$ of SMGs to be 
$\sim 10^{-5}\;h_{70}^{3}$~Mpc$^{-3}$ from the observed area covered by the SCLS  
and the estimated redshift range $1<z<5$.
  The duty cycle of SMGs can be constrained directly from the gas depletion time scale.
 \citet{GenzelR_03a} showed that such a luminous galaxy (with a SFR
$\sim 500$\mpy) would use up its gas in approximately $4\times10^8$~yr
and make $\sim 2\times10^{11}$~\msun\ of stars.  \citet{TeczaM_04a}
estimated an age for J14011$+$0252 of $\geq$200 Myrs from the strength
of its Balmer break.  
Using a SMG time scale of $4\times10^8$~yr, and assuming that SMGs are evenly distributed
over 1$\leq$z$\leq$5 ($\simeq 4.5$~Gyr), this would imply that SMGs are `on'
10\%\ of the time or have a duty cycle $r$ of 0.1.
Thus, the co-moving density is  at least
$n\ga1.3\times10^{-4}\;h_{70}^3\;\left (\frac{r}{0.1}\right
)^{-1}$~Mpc$^{-3}$ \citep{GenzelR_03a}.
Independently, \cite{ChapmanS_05a} estimated a raw co-moving
density $n$ of  $z\simeq 2.5$ SMGs above L$_{FIR}=10^{12.5}$ of $n\sim 10^{-5}
\;h_{70}^{3}$~Mpc$^{-3}$ using measured redshifts
of radio identified SMGs. \cite{ChapmanS_05a} modeled the redshift distribution of SMGs with a Gaussian
distribution of width $\sigma_z\simeq 1.3$ (covering $\sim$~1~Gyr), and assumed
a time scale of $10^8$~yr,  yielding a similar duty cycle $r$ of 0.1.
Thus, the true co-moving space density estimated by two groups,
$n=$1--3$\times10^{-4}\;h_{70}^3\;\left (\frac{r}{0.1}\right
)^{-1}$~Mpc$^{-3}$.

\begin{figure*}
\centerline{\includegraphics[width=145mm]{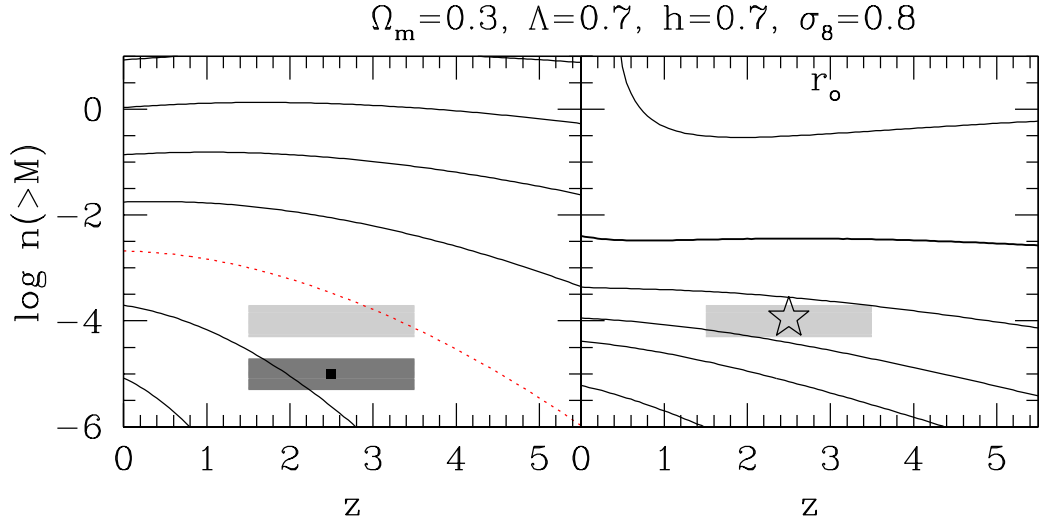} }
\caption{{\it Left}: Co-moving number density of sources vs. redshift.
Lines of constant halo mass are shown by the contours for $\log M_h=14,\;13,\;11,\;10,\;9$,
from bottom to top. The filled square shows the observed number density of SMGs, 
$n\simeq1\times10^{-5}$~\hMpcc, with the uncertainty and redshift coverage represented by
the dark shaded area. {\it Right}: Same as left, but the contours represent lines
of constant clustering amplitude $r_0=15,\;10,\;8,\;6,\;4,\;2$~Mpc (bottom to top). The correlation length
of SMGs \citep[$r_0\simeq 7$~\hMpc][]{BlainA_04a} is shown by the star.
The halo mass inferred from the clustering is $\sim10^{12.5}$~\msun. Given that the observed
$n$ (left panel, solid square) is lower than the one inferred from the clustering (star) by a factor
of $\sim10$,  the duty cycle of SMGs is about $\sim0.1$. The number density of SMGs corrected
for the duty cycle is shown by the light shaded area in both panels. Both panels were produced
using the Press-Schechter formalism of \citet{MoH_02a}.
}
\label{fig:mowhite}
\end{figure*}

Another way to estimate the duty cycle of SMGs is to use  their
clustering strength \citep[as pointed out by][]{ChapmanS_05a}) and Fig.~\ref{fig:mowhite}.
 Both panels show the comoving number density of sources vs. redshift.
In the left panel, lines of constant halo mass $\log M_h=14,\;13,\;11,\;10,\;9$ are shown.
In the right panel, lines of constant clustering amplitude $r_0=15,\;10,\;8,\;6,\;4,\;2$~Mpc  (from bottom to top) are shown.
  \citet{BlainA_04a} estimated the SMG auto-correlation length $r_0$ to be $r_0=7\;h_{70}^{-1}$~Mpc
  (represented by the star), from which one would infer a halo co-moving abundance density of
$n\simeq1\times10^{-4}\;h_{70}^3$~Mpc$^{-3}$ (light shaded area). This is $\approx$10   times larger
than the observed number density ($n\sim 1\times 10^{-5}$~Mpc$^{-3}$, dark shaded area).
These two numbers can be reconciled if SMGs are short lived, with a
``duty cycle'' of the order of 0.1.
Nonetheless, the several methods of estimating the duty cycle of SMGs
agree to within a factor of a few.

\subsection{Consequences of SMGs properties}

From the observed properties of SMGs (summarized in Table~\ref{table:submm:observations}),
in this section, we  derive the co-moving baryonic
density and co-moving metal densities in SMGs (summarized in Table~\ref{table:SMG:observations}).

The dynamical masses of SMGs cover the range $2$---$3\times 10^{11}\;h_{70}^{-1}\msun$ (Table~\ref{table:submm:observations}) and
assuming that this mass is entirely baryonic, the co-moving baryonic
density in submm galaxies is observed to be:

\begin{equation}
\rho_{b,\rm obs}\sim 2.7\times10^7\;h_{70}^2(\frac{r}{0.1})^{-1}~\rhosun\;.\label{baryons:genzel}
\end{equation}

The observed mean gas mass $M_{\rm gas}$ of  $\simeq 4.0\times
10^{10}$~\msun\ corresponds to a gas fraction $f_g=20$\%\ and
implies a gas (molecular) co-moving density of  $\rho_{\rm gas,
SMG}=5.2\times10^6\;h_{70}^2\rhosun$.

Naturally, the ISM of the SMGs contains also an unknown amount of neutral gas (\HI).
  Since we do not know the neutral to molecular ratios, it is
difficult to make a robust estimate of the total potential reservoir
of gaseous material.  In the local Universe, \cite{KeresD_03a} have estimated the
co-moving mass contribution of the cold molecular gas (\Htwo) and compared it to the atomic phase (\HI)
of the interstellar media of galaxies.  They find  an average ratio
of \Htwo\ to \HI\ of about 0.5 to 0.65. The molecular-to-neutral ratio
increases with increasing \Htwo\  mass and is about 2 for galaxies with
M$_{\Htwo}\sim10^9$~\msun. At very high infrared luminosities, similar
to what is found for the SMGs, galaxies show \Htwo\ to \HI\ mass ratios of
$\sim 4$ to 20 \citep{MirabelI_89a}. Using these local results,
in order to provide a more robust estimate of the total metal content,
we  assume that
(i) there is as much HI as molecular gas, which is likely to be an upper
limit given the local ratios \citep{MirabelI_89a}, and 
that (ii) the metallicity of SMGs is
on average $<Z>\sim 1.9\;Z_\odot$, the metallicity
of SMM14011$+$1152.
We  find that the metal co-moving density in SMGs is:
\begin{equation}
\rho_{Z,\rm obs}\la
3.6\times10^5\;h_{70}^2\frac{f_g}{40\%}\left (\frac{r}{0.1}\right )^{-1}\frac{<Z>}{1.9}~\rhosun\;,\label{metals:genzel}
\end{equation}
or $\la~9$\%\ of the metal budget.

We also note that our assumption that there is as much HI as molecular gas implies
that SMGs can  only add $\approx\ 3.5$\%\ to the neutral gas content of the universe $\Omega_{\rm DLA}$, and thus SMGs
can not harbor the dusty DLAs of \citet{VladiloG_04a}. The two are distinct populations.

It is absolutely necessary to set a firm upper limit to the contribution
of SMGs to the cosmic metal density.  This is especially true in light
of the  claim by D03 that the metal content
of SMGs is sufficient to solve the missing metals problem (more on
this in \S~\ref{section:dunne}).  Eq.~\ref{metals:genzel} is a hard upper limit from
two independent lines of arguments.
First, it is the maximum gas-mass fraction allowed by the metallicity.
Indeed, high metallicity and large gas fractions are mutually exclusive 
in chemical evolution models. \citet{EdmundsM_90a} showed that
when the gas fraction is much larger than 50\%\, solar metallicity cannot be reached.
In addition,  the chemical evolution models of \citet{EdmundsM_98a} (used by D03),
showed that the stellar contribution to the metal mass budget of SMG
is small (approximately 1/4 of Eq.~\ref{metals:genzel})
given the lower metallicity of the stellar component compared to the gas component (see section~3).
Second,   our mean metallicity of $<Z>=1.9$ is large.
If we assume $Z=Z_\odot$ and treat both the gas {\it and} the stellar component equally,
 the SMG contribution to the metal budget  would remain the same. In this case,
  Eq.~\ref{metals:genzel} is a strong upper limit given that 
   100\%\ of the dynamical mass $M_{\rm dyn}$ is in baryons
   (i.e.  neglecting any contribution from dark matter to the dynamical mass).

Thus, we conclude that Eq.~\ref{metals:genzel} is a strong upper limit given
our assumptions. 
SMGs with $S_{850}\ga 3$mJy  can not contribute  more than $\sim$9\%\ of the $z=2$ metal budget.

\begin{table*}
\caption{Baryons and metal cosmic densities in SMGs from this paper.\label{table:SMG:observations}}
\begin{tabular}{lclrlcccr}
\hline
		&  $\rho(\rhosun)$ 	& $\rho/\rho_c$ 	 	&  $\rho/\rho_b$  (\%)	 & Note &	$Z/Z_\odot$ &
		$\rho_Z$(\rhosun) 	& $\rho_Z/\rho_c$  & $\rho_Z/\rho_{Z,\rm tot}$ (\%) 	\\
Baryons	&  $ 5.98\times10^9\;h_{70}^{1}$ &  $0.044\;h_{70}^{-2}$	&	100   	 \\
\hline
Stars $ 2<z<4$	&  $1.21\times10^{8}\;h_{70}^0$  &  $0.000890\;h_{70}^{-2}$ 	& 2.02	& $\int$ SFR~\tablenotemark{a}	& 1.25~\tablenotemark{b}
& $4.00\times10^{6}$	& $2.94\times10^{-5}$	& 100	 \\
\hline \hline
SMGs $>3$mJy: \\
\hline 
SMGs Baryons	&  $2.73\times10^{7}\;h_{70}^2$ 	&  $0.000201\;h_{70}^0$ 	& 0.46	& 	& 	& 	& 	&   \\

SMGs ISM (H2)	&  $5.16\times10^{6}\;h_{70}^2$ 	&  $0.000038\;h_{70}^0$ 	& 0.09	& $f_g=19$\%	& 	&  	&  		  \\
$-->$ ISM (H2+HI)&  $1.03\times10^{7}\;h_{70}^2$ 	&  $0.000076\;h_{70}^0$ 	& 0.17	& $f_g=38$\%	& 1.9	& $3.63\times10^{5}$	& $2.67\times10^{-6}$	& 9.1	  \\
\hline
\end{tabular}
\tablenotetext{a}{Cosmic stellar density calculated from the integrated
SFH taking into account a recycled fraction of $R=0.28$. $f_g$ is the
gas fraction.} 
\tablenotetext{b}{Averaged yield ($=1/42=1.25\;Z_\odot$) for type II SN with
$m>10$~\msun\  \citep{MadauP_96a}.}
\end{table*}

\begin{table*}
\caption{Results from the dust mass function (D03).
Numbers in bold are taken from D03 and corrected for our
cosmology.\label{table:dunne}}
\begin{tabular}{lclrlcccr}
\hline
		&  $\rho(\rhosun)$ 	& $\rho/\rho_c$ 	 	&  $\rho/\rho_b$ (\%) 	 & Note &	$Z/Z_\odot$ &
		$\rho_Z$(\rhosun) 	& $\rho_Z/\rho_c$  & $\rho_Z/\rho_{Z,\rm tot}$(\%) 	\\
Baryons	&  $ 5.98\times10^9\;h_{70}^{1}$ &  $0.044\;h_{70}^{-2}$	&	100    	 \\
\hline
Stars $ 2<z<4$	&  $1.21\times10^{8}\;h_{70}^0$  &  $0.000890\;h_{70}^{-2}$ 	& 2.02	& $\int$ SFR~\tablenotemark{a}	&
1.25~\tablenotemark{b}	& $4.00\times10^{6}$	& $2.94\times10^{-5}$	& 100 \\
\hline \hline
SMGs dust (DMF)	&  $\mathbf{4.39\times10^{5}\;h_{70}^1}$ 	&  $0.000003\;h_{70}^{-1}$ 	& 0.01	&     \\
SMG baryons	&  $\mathbf{7.65\times10^{7}\;h_{70}^1}$ 	&  $0.000562\;h_{70}^{-1}$ 	& 1.28	&   	&  	&  	&  	&  \\
 --  ISM       &  $3.82\times10^{7}\;h_{70}^1$         &  $0.000281\;h_{70}^{-1}$      & 0.64  & $\eta=0.4$    & 1.52  & $1.10\times10^{6}$    & $8.07\times10^{-6}$   & 27.4  \\
 -- Stars     &  $3.82\times10^{7}\;h_{70}^1$         &  $0.000281\;h_{70}^{-1}$      & 0.64  &     & 0.33  & $2.41\times10^{5}$    & $1.78\times10^{-6}$   & 6.0     \\
\hline 
SMG bar. ( $>3$mJy):	&  $4.47\times10^{7}\;h_{70}^1$ 	&  $0.000329\;h_{70}^{-1}$ 	& 0.75	&     &       &       &       &       \\
 --  ISM       &  $2.24\times10^{7}\;h_{70}^1$         &  $0.000164\;h_{70}^{-1}$      & 0.37  &     &       & $6.42\times10^{5}$    & $4.72\times10^{-6}$   & 16.0    \\
 -- Stars     &  $\mathbf{2.24\times10^{7}\;h_{70}^1}$ &  $0.000164\;h_{70}^{-1}$     & 0.37  &     & $->$0.33  & $\mathbf{1.41\times10^{5}}$    & $1.04\times10^{-6}$   & 3.5   \\
		\hline	
			
\tablenotetext{a}{Cosmic stellar density calculated from the integrated
SFH taking into account a recycled fraction of $R=0.28$. $\eta=0.4$
is the fraction of the ISM metals that are assumed to be
locked onto dust grains.}
\tablenotetext{b}{Averaged yield ($=1/42=1.25\;Z_\odot$) for type II SN with $m>10$~\msun\ \citep{MadauP_96a}.}
\end{tabular}
\end{table*}

\section{The dust mass function and its implications}
\label{section:dunne}

D03 took a very different approach. They used submm data from published
deep blank field SCUBA surveys and constructed the dust mass function
(DMF) of high redshift galaxies.  From the DMF, they inferred,
using chemical models, the co-moving density of metals and baryons
associated with the ISM of galaxies.  They assumed a dust temperature
of $\sim25$~K from \citet{PeiY_99a}.  Because the dust mass is a strong
function of the dust temperature, a lower (higher) dust
temperature 20~K (30~K) will increase (decrease) their mass estimates by a
factor of about  two (D03).  D03 argued that most of the dust mass in SMGs would be
a low temperature ($T_d\sim 20$~K) component. We note that the recent observations
of \citet{ChapmanS_05a} favor higher dust temperature ($T_d\simeq35$~K).
D03 also used the dust mass opacity measured at 125$\mu$m and extrapolated
it to submm wavelengths using a $\lambda^{-\beta}$ dependence with
$\beta=2$ \citep[from their local survey][]{DunneL_01a}.  We now discuss
their results, which are summarized in Table~\ref{table:dunne}.

\subsection{DMF all}

D03 find that the DMF function is well described by a Schechter
function  with the parameters $M_d^*=4.7\times10^8$\msun,
$\phi_d^*=8.9\times10^{-4}$~Mpc$^{-3}$, and $\alpha=-1.08$ in a $\Lambda$
cosmology.  The integral of the DMF gives the co-moving density of dust
$\rho_d=\Gamma(2+\alpha)M_d^*\phi_d^*=4.39\times10^5\;\rhosun$~\footnote{D03
used a $\Omega_M=1$ cosmology throughout their paper.  In the
remainder of this paper, we used their DMF  in a $\Omega_M=0.3$,
$\Lambda=0.7$ cosmology.  This lowers their cosmic densities
($\rho_{b,\rm DMF}=1.3\times10^8\;h_{75}~\rhosun$, $\rho_{Z,\rm
DMF}=1.9\times10^6\;h_{75}\rhosun$) by a  factor  1.7.
The global factor between our cosmology and D03 is 1.8 including the
change from $h_{75}$ to $h_{70}$.}.  From $\rho_d$,
they assumed that 40\%\ ($\eta=0.4$) of the ISM metals are locked into
dust grains, yielding a  metal density:

\begin{equation}
\rho_{Z,\rm DMF}=1.1\times10^6\;h_{70}\frac{0.4}{\eta}\;\rhosun\;,\label{metals:dunne}
\end{equation}
i.e., about 27\%\ of the cosmic metal budget.  The
apparent discrepancy between that number and the original conclusion of
D03, namely that SMGs contain $\ga70$\%\ of the cosmic metal density, is due
to the different cosmology assumed.  

A strong lower limit comes from $\eta=1$, i.e. assuming 100\%\ of the metals are
locked onto the submm emitting dust grains. In that case, at least 11\%\
of the metals are in the ISM of SMGs. Note that (i) this does {\it not}
include the metals in stars, and (ii) it will not depend on the chemical evolution models discussed below.

D03  used a ``closed-box'' chemical evolution model \citep{EdmundsM_98a}
to convert the dust content into a total baryonic content. They  find
that SMGs contribute  a baryonic co-moving density  of about:
\begin{equation}
\rho_{b,\rm DMF}\simeq 7.2\times10^7\;h_{70}\;\rhosun. \label{baryons:dunne}
\end{equation}

 In these chemical evolution models, they have assumed that the submm sources are observed
at their maximum dust mass, i.e., at the peak of the dust mass to baryonic
mass ratio. Thus their baryonic mass density is a strong lower limit.
This peak occurs at a gas fraction of roughly 50\%.

As a side note, the inferred metallicity from D03 is close to the observed metallicity of SMGs.
From the metallicity density (Eq.~\ref{metals:dunne}) and
the baryonic density (Eq.~\ref{baryons:dunne}, obtained from the chemical models),
  the mean metallicity of the ISM
is $\rho_{Z,DMF}/(\rho_{b,\rm DMF}\cdot f_g)$ or about $\sim
1.51\;Z_\odot$, assuming that all the metals are in the ISM and a
50\%\ gas fraction $f_g$.  This mean metallicity is very close to
the observed metallicity of one SMG (J14011) discussed in
section~\ref{section:SMG:observations}.

\subsection{DMF bright}

In order to compare this prediction to the observed properties of
SMGs discussed in section~\ref{section:SMG:observations}, one needs to
compare the cosmic baryonic and metal densities of flux selected
SMGs to the DMF with a similar flux limit.  D03 integrated the
DMF down to $S_{850}>3$~mJy, but quoted only the stellar mass
density ($\rho_{\rm star}=2.24\times10^7\;h_{70}$~\rhosun;
half of the baryons) and stellar metal density ($\rho_{Z,\rm
star}=1.4\times10^5\;h_{70}$~\rhosun)~\footnote{We
again converted their number ($\rho_{b,\rm
DMF}=2\times\rho_\star=2\times3.8\times10^7\;h_{70}~\rhosun$) in a
$\Omega_M=1$ to a $\Lambda$ cosmology.}.  Thus, the cosmic baryon
density for bright (with $S_{850}>3$~mJy) submm sources of (twice the
stellar density):

\begin{eqnarray}
\rho_{b,\rm DMF,3mJy}&\simeq& 4.2\times10^7\;h_{70}\rhosun
\label{baryons:dunne:bright}
\end{eqnarray}
i.e., $\sim 0.6$ times the number quoted in Eq.~\ref{baryons:dunne},
providing the completeness factor.
We then scale the total metal cosmic density (Eq.~\ref{metals:dunne})
by this $0.6$ factor to infer the metal cosmic density in sources with
$S_{850}>3$mJy:
\begin{eqnarray}
\rho_{Z,\rm DMF,3mJy}&\simeq& 6.4\times10^5\;h_{70}\rhosun\;,\label{metals:dunne:bright}
\end{eqnarray}
or about $\sim$16\%\ of the estimated cosmic metal density
(Eq.~\ref{eq:metals:madau}).

If one compares the  baryonic and metal densities predicted from the DMF
(Eqs.~\ref{baryons:dunne:bright}, \ref{metals:dunne:bright}) with 
the observed baryonic and metal densities (Eqs.~\ref{baryons:genzel}, \ref{metals:genzel}),
 we conclude that the predictions  from the DMF
(in a $\Lambda$  cosmology) were over-estimated by a factor of at least 2.  This factor can be easily
accounted for if one uses a higher dust temperature ($T_d\simeq 35$~K), as indicated by
the observations of \citet{ChapmanS_05a}.

\section{Summary \&\ Discussion}
\label{section:conclusions}

SMGs are gas- (gas fraction 20--50\%) and metal-rich galaxies ($Z/Z_\odot\ga1$).
Therefore, they are potentially  good candidates for harboring
the missing metals.  From the observed gas fraction $\sim 40$\%, dynamical mass $M_{\rm dyn}$  and 
 mean metallicity of $>Z_\odot$
\citep[supported by][]{TeczaM_04a}  of 7 SMGs brighter than $S_{850}>3$~mJy,
we show that 

\begin{itemize}
\item Based on the dynamical masses of SMGs an
assuming a 100\%\ baryon fraction with $<Z>=Z_\odot$,
 SMGs can not contribute more than 9\%\ of the expected cosmic metal density;
\item Based the observed high gas fractions and observed high ISM metallicities $Z>>Z_\odot$,
 SMGs can not contribute more than 9\%\ of the expected cosmic metal density;
\item the total contribution of  SMGs, correcting for incompleteness (section~3.2),
is  $\frac{1}{0.6}$ times the contribution of SMGs brighter than $3$~mJy,
or $\la15$\%;
\item our results imply that SMGs can  only add $\approx\ 3.5$\%\ to $\Omega_{\rm DLA}$.
Thus SMGs cannot harbor the dusty DLAs of \citet{VladiloG_04a}.
\end{itemize}

 Early estimates of the contribution of the SMGs to the metal budget from the DMF were
overestimated. The discrepancy however is mainly
due to the assumed cosmology  and to the low dust temperature used by D03 ($T_d\simeq 25$~K).
We do agree with the conclusion of D03 that SMGs make a significant
contribution to the cosmic metal budget, just not enough to solve the
``missing metals problem''.

We are still far from closing the metal budget, however. 
In addition to $\la$9\%\ of the metals that are in SMGs,
5\%\ are in $z\sim3$ LBGs \citep{PettiniM_03b},
 $\sim15$--20\%\ in $z\sim2.2$   galaxies (see paper~II),
8\%\ in the forest (but see paper~III), and
5\%\ in DLAs~\footnote{Dusty DLAs  \citep{VladiloG_04a}, missed in current spectroscopic
DLA surveys \citep[current DLA samples show small molecular and dust contents,
e.g.][]{EllisonS_01c,MurphyM_04c} could amount to an additional
17\% (paper~III). These would be in a separate population from SMGs
given that the amount of \HI\ in SMGs is less than 3--4\%\ of $\Omega_{\rm DLA}$.}.
 Taking our results on SMGs at face-value, and ignoring the issue of
double-counting, roughly 50\%\ of the metals have been accounted for (see also paper~III).

We are exploring two main avenues in trying to close the
missing metals problem.  Following
\citet{PettiniM_03b},   either another population of galaxies
  has not yet been accounted for or there is a significant
reservoir of metals in the IGM that has not been detected.
A  substantial fraction of the missing
metals may be hidden in a very hot, collisionally ionized gas.  Based on
simple order-of-magnitude calculations,  in paper~III, we will discuss the possibility that the remaining
missing metals could have been ejected from small galaxies via galactic
outflows into the IGM in a hot (T$>10^6$~K) that is difficult to detect
using observed properties of local galaxies.

\section*{Acknowledgments}
We thank T. Greve for providing the $S_{850}$ fluxes of the SMGs;
A. Baker, L. Tacconi, D. Lutz and M. Pettini for helpful comments and discussions.
We thank the referee for his/her comments that improved the quality of the paper.

\bsp_small

\label{lastpage}

\end{document}